\documentclass[twocolumn,showpacs,preprintnumbers,amsmath,amssymb]{revtex4}

\usepackage{graphicx}
\usepackage{dcolumn}
\usepackage{bm}

\begin{document}

\preprint{APS/123-QED}

\title{Quantum gapped state in a spin-1/2 distorted honeycomb-based lattice with frustration}

\author{Hironori Yamaguchi$^{1}$, Satoshi Morota$^{1}$, Takanori Kida$^{2}$, Seiya Shimono$^{3}$, Koji Araki$^{4}$, Yoshiki Iwasaki$^{5}$, Yuko Hosokoshi$^{1}$, and Masayuki Hagiwara$^{2}$}

\affiliation{
$^1$Department of Physics, Osaka Metropolitan University, Osaka 599-8531, Japan\\
$^2$Center for Advanced High Magnetic Field Science (AHMF), Graduate School of Science, Osaka University, Osaka 560-0043, Japan\\
$^3$Department of Materials Science and Engineering, National Defense Academy, Kanagawa 239-8686, Japan\\
$^4$Department of Applied Physics, National Defense Academy, Kanagawa 239-8686, Japan\\
$^5$Department of Physics, College of Humanities and Sciences, Nihon University, Tokyo 156-8550, Japan
}


Second institution and/or address\\
This line break forced

\date{\today}

\begin{abstract}
We successfully synthesized ($p$-Py-V)[Cu(hfac)$_2$], a verdazyl-based complex.
Molecular orbital calculations revealed five types of intermolecular interactions between the radical spins and two types of intramolecular interactions between the radical and the Cu spins, resulting in a spin-1/2 distorted honeycomb-based lattice.
Additionally, competing ferromagnetic and antiferromagnetic (AF) interactions induce frustration.
The magnetization curve displayed a multistage increase, including a zero-field energy gap. 
Considering the stronger AF interactions that form dimers and tetramers, the magnetic susceptibility and magnetization curves were qualitatively explained.
These findings demonstrated that the quantum state, based on the dominant AF interactions, was stabilized due to the effects of frustration in the lattice.
Hence, the exchange interactions forming two-dimensional couplings decoupled, reducing energy loss caused by frustration and leading to frustration-induced dimensional reduction.
\end{abstract}


\maketitle
\section{INTRODUCTION}
Honeycomb lattices have captured significant interest in the field of condensed-matter physics due to their intriguing topological properties. 
The honeycomb lattice antiferromagnet exhibits a bipartite structure, eliminating magnetic frustration caused by nearest-neighbor exchange interactions. 
Notably, the honeycomb lattice has a minimal coordination number among two-dimensional (2D) systems, resulting in increased quantum fluctuations.
Although the ground state of this lattice demonstrates long-range order, the presence of strong quantum fluctuations diminishes the magnetic moment per site, making the ordered state vulnerable~\cite{honeycombM1,honeycombM2,honeycombM3,honeycombM5}.
Consequently, even minor perturbations such as lattice distortion~\cite{dimer1, dimer2, dimer3} and randomness~\cite{uematsu} can easily destabilize the ordered state.

The combination of strong quantum fluctuations and frustration in a honeycomb lattice offers an excellent platform for investigating quantum many-body phenomena.
Frustration is introduced into the honeycomb lattice through competing distant exchange interactions.
Theoretical studies on frustrated honeycomb lattices predict a variety of competing phases in both classical and quantum regimes~\cite{f_hone1,f_hone2,f_hone3,f_hone4}. 
In certain parameter regions, magnetically ordered states are destabilized through quantum fluctuations, resulting in quantum phases such as a gapped quantum spin liquid and a plaquette valence bond crystal.
For the distorted honeycomb lattice, which is the focus of this study, frustration is expected to induce dimensional reduction.
The exchange couplings forming the lattice are partially decoupled to diminish competing interactions and minimize the ground-state energy. 
Several frustrated 2D systems have been reported to exhibit 1D quantum behavior, demonstrating dimensional reduction due to the effect of frustration~\cite{tri1,tri2,tri3, PF6, Zn_gap}.


Form-designed radicals with diverse molecular structures have proven to be highly effective in creating honeycomb lattices and realizing a  variety of spin-1/2 honeycomb-based lattices, ranging from one-dimensional honeycomb chains to three-dimensional honeycomb network~\cite{2Cl6FV, iwase, random, 3D, Zn_honeycomb}.
Our previous studies on spin-1/2 distorted honeycomb lattices composed of form-designed radicals have revealed the emergence of a gapped singlet state~\cite{3D} and valence bond glass induced by randomness~\cite{random}.
Furthermore, the presence of alternating spin density distribution, resulting from $\pi$-conjugated systems, can induce ferromagnetic intermolecular exchange interactions depending on the overlapping patterns of molecular orbitals~\cite{PF6, Zn_gap, peierls}. 
Recently, we have achieved an expansion of our spin arrangement design by combining it with transition metals.
By converting form-designed radicals into ligand structures, we introduced metal-radical couplings and magnetic anisotropy into our spin model designs. 
The complexes formed with 3$d$ transition metals exhibited spin systems composed of intermolecular $\pi-\pi$ stacking and intramolecular $\pi-d$ couplings~\cite{morotaMn, 2DCo}, resulting in the distorted honeycomb-based lattice employed in this study.

Herein, we successfully synthesized ($p$-Py-V)[Cu(hfac)$_2$] ($p$-Py-V = 3-(4-pyridinyl)-1,5-diphenylverdazyl, hfac = 1,1,1,5,5,5-hexafluoro-2,4-pentanedione) that is a verdazyl-Cu complex. 
Molecular orbital (MO) calculations revealed five types of intermolecular interactions between the radical spins and two types of intramolecular interaction between the radical and the Cu spins, resulting in a spin-1/2 distorted honeycomb-based lattice.
The competition between ferromagnetic and antiferromagnetic (AF) interactions introduced frustration.
The magnetization curve displayed a multistage increase, including a zero-field energy gap. 
The behavior of the magnetic susceptibility and magnetization curves can be largely explained by the AF interactions forming dimers and tetramers, indicating the stabilization of the quantum state based on the dominant AF interactions within the lattice.
Moreover, we demonstrate that frustration plays a crucial role in stabilizing the quantum-gapped state.
These findings suggest that the exchange interactions forming the two-dimensional lattice undergo decoupling to reduce the energy loss caused by frustration, resulting in a frustration-induced dimensional reduction.

\section{EXPERIMENTAL}
We synthesized $p$-Py-V via the conventional procedure for producing the verdazyl radical~\cite{verd}.
A solution of Cu(hfac)$_2$$\cdot$2H$_2$O (205.5 mg, 0.45 mmol) in 2 ml ethanol and 10 ml of heptane was refluxed at 60 $^\circ$C. 
A solution of $p$-Py-V (251.5 mg, 0.80 mmol) in 4 ml of CH$_2$Cl$_2$ was slowly added and stirred for 1 h. 
After the mixed solution was cooled to room temperature, a dark-brown crystalline solid of ($p$-Py-V)[Cu(hfac)$_2$] was separated by filtration and washed with heptane.
Single crystals were obtained via recrystallization from a mixed solvent of CH$_2$Cl$_3$ and $n$-heptane at 10 $^\circ$C.

The X-ray intensity data were collected using a Rigaku XtaLAB Synergy-S instrument.
The crystal structures was determined using a direct method using SIR2004~\cite{SIR2004} and refined using the SHELXL97 crystal structure refinement program~\cite{SHELX-97}.
Anisotropic and isotropic thermal parameters were employed for non-hydrogen and hydrogen atoms, respectively, during the structure refinement. 
The hydrogen atoms were positioned at their calculated ideal positions.
Magnetization measurements were conducted using a commercial SQUID magnetometer (MPMS-XL, Quantum Design).
The experimental results were corrected for the diamagnetic contribution, which are determined based on the numerical analysis to be described and confirmed to be close to that calculated by Pascal's method.
High-field magnetization in pulsed magnetic fields was measured using a non-destructive pulse magnet at AHMF, Osaka University.
Specific heat measurements were performed using a commercial calorimeter (PPMS, Quantum Design) employing a thermal relaxation method.
All the experiments utilized small, randomly oriented single crystals.

Molecular orbital (MO) calculations were performed using the UB3LYP method as broken-symmetry hybrid density functional theory calculations with a basis set of 6-31G(d, p). 
All calculations were performed using the GAUSSIAN09 software package.
The convergence criterion was set at 10$^{-8}$ hartrees.
We employed a conventional evaluation scheme to estimate the intermolecular exchange interactions in the molecular pairs~\cite{MOcal}. 

The quantum Monte Carlo (QMC) code is based on the directed loop algorithm in the stochastic series expansion representation~\cite{QMC2}. 
The calculations was performed for $N$ = 1152 under the periodic boundary condition, where $N$ denotes the system size.
It was confirmed that there is no significant size-dependent effect.
All calculations were carried out using the ALPS application~\cite{ALPS,ALPS3}.

\section{RESULTS}
\subsection{Crystal structure and spin model}
The crystallographic parameters for ($p$-Py-V)[Cu(hfac)$_2$] are listed in Table I.
The crystals consist of two distinct molecules, as depicted in Fig. 1(a).
For each molecule, a verdazyl radical, $p$-Py-V, and a Cu$^{2+}$ ion possess spin value of 1/2.
The Cu$^{2+}$ ion is coordinated by the $p$-Py-V ligand and four O atoms in hfac, resulting in a 5-coordinate environment.
The bond lengths and angles of the Cu atoms are listed in Table II.
MO calculation revealed that approximately 60${\%}$ of the total spin density is localized on the central ring consisting of four N atoms, and the phenyl rings directly attached to the central N atom contribute approximately 15-18${\%}$ of the spin density each.
The pyridine ring that lacks a direct connection to the N atom accounts for less than 8${\%}$ of the spin density.
Dominant exchange interactions were determined through MO calculations. 
In the intermolecular case, five primary exchange interactions were identified among the radicals, as depicted in Fig. 1(b).
Their values are evaluated as $J_{\rm{V1}}/k_{\rm{B}}$ = 21.4 K, $J_{\rm{V2}}/k_{\rm{B}}$ = 3.4 K, $J_{\rm{V3}}/k_{\rm{B}}$ = 3.2 K, $J_{\rm{V4}}/k_{\rm{B}}$ = 2.9 K, and $J_{\rm{V5}}/k_{\rm{B}}$ = $-$7.2 K, defined within the Heisenberg spin Hamiltonian, given by $\mathcal {H} = J_{n}{\sum^{}_{<i,j>}}\textbf{{\textit S}}_{i}{\cdot}\textbf{{\textit S}}_{j}$, where $\sum_{<i,j>}$ denotes the sum over neighboring spin pairs.
While two molecules associated with $J_{\rm{V2}}$ are crystallographically independent, the molecular pairs associated with the other interactions are related by inversion symmetry.
These five interactions formed a spin-1/2 distorted honeycomb lattice in the $ab$-plane, as shown in Fig. 1(b).
Frustration is induced by four AF interactions and one ferromagnetic interaction.
Additionally, the presence of nonmagnetic hfac moieties between the 2D structures enhances the two-dimensionality of the spin lattice, as shown in Fig. 1(c).
In the intramolecular case, AF exchange interactions between the spins on the radicals and the Cu atoms in both molecules were evaluated.
The magnitudes of these couplings for M1 and M2 were evaluated as $J_{\rm{Cu1}}/k_{\rm{B}}$ = $41$ K and $J_{\rm{Cu2}}/k_{\rm{B}}$ = $36$ K, respectively.
Notably, MO calculations tend to overestimate the intramolecular interactions between verdazyl radicals and transition metals~\cite{morotaMn}; therefore, the actual values of $J_{\rm{Cu1}}$ and $J_{\rm{Cu2}}$ are expected to be smaller than the MO evaluations.
Consequently, each spin site on the honeycomb lattice formed by $J_{\rm{V1}}$-$J_{\rm{V5}}$ is connected to the Cu spin via $J_{\rm{Cu1}}$ or $J_{\rm{Cu2}}$, establishing a spin-1/2 distorted honeycomb-based lattice, as depicted in Fig. 2.

\begin{table}
\caption{Crystallographic data for ($p$-Py-V)[Cu(hfac)$_2$].}
\label{t1}
\begin{center}
\begin{tabular}{cc}
\hline
\hline 
Formula & C$_{29}$H$_{18}$CuF$_{12}$N$_{5}$O$_{4}$\\
Crystal system & Triclinic \\
Space group & $P\bar{\rm{1}}$ \\
Temperature (K) & 100 \\
$a$ $(\rm{\AA})$ & 10.7900(3) \\
$b$ $(\rm{\AA})$ & 14.6632(4)  \\
$c$ $(\rm{\AA})$ & 19.2759(6)  \\
$\alpha$ (degrees) & 92.602(2) \\
$\beta$ (degrees) & 99.147(2)\\
$\gamma$ (degrees) & 91.583(2)\\
$V$ ($\rm{\AA}^3$) & 3005.99(15) \\
$Z$ & 4 \\
$D_{\rm{calc}}$ (g cm$^{-3}$) & 1.750\\
Total reflections & 7511 \\
Reflection used & 6548 \\
Parameters refined & 919 \\
$R$ [$I>2\sigma(I)$] & 0.0567  \\
$R_w$ [$I>2\sigma(I)$] & 0.1576 \\
Goodness of fit & 1.039 \\
CCDC & 2280049\\
\hline
\hline
\end{tabular}
\end{center}
\end{table}

\begin{table}
\caption{Bond distances [$\rm{\AA}$] and angles [$^{\circ}$] related to two crystallographically independent Cu atoms for ($p$-Py-V)[Cu(hfac)$_2$].}
\label{t1}
\begin{center}
\begin{tabular}{cc@{\hspace{1cm}}cc}
\hline
\multicolumn{2}{c}{M$\bf{1}$} & \multicolumn{2}{c}{M$\bf{2}$}\\
\hline
Cu1--N1 & 2.03 & Cu2--N2 & 2.01\\
Cu1--O1 & 1.95 & Cu2--O5 & 1.95\\
Cu1--O2 & 1.98 & Cu2--O6 & 2.22\\
Cu1--O3 & 2.22 & Cu2--O7 & 1.98\\
Cu1--O4 & 1.95 & Cu2--O8 & 1.95\\
 & & & \\
N1--Cu1--O1 & 92.2 & N2--Cu2--O5 & 91.1\\
O1--Cu1--O2 & 90.5 & O5--Cu2--O7 & 87.1\\
O2--Cu1--O4 & 85.3 & O7--Cu2--O8 & 91.0\\
O4--Cu1--N1 & 91.1 & O8--Cu2--N2 & 92.2\\
O1--Cu1--O3 & 97.3 & O5--Cu2--O6 & 88.1\\
O3--Cu1--O4 & 88.9 & O6--Cu2--O8 & 87.3\\
O4--Cu1--O1 & 172.3 & O8--Cu2--O5 & 175.0\\
N1--Cu1--O3 & 100.4 & N2--Cu2--O6 & 109.4\\
O3--Cu1--O2 & 86.8 & O6--Cu2--O7 & 89.7\\
O2--Cu1--N1 & 171.9 & O7--Cu2--N2 & 160.8\\
\hline
\end{tabular}
\end{center}
\end{table}

\begin{figure*}[t]
\begin{center}
\includegraphics[width=38pc]{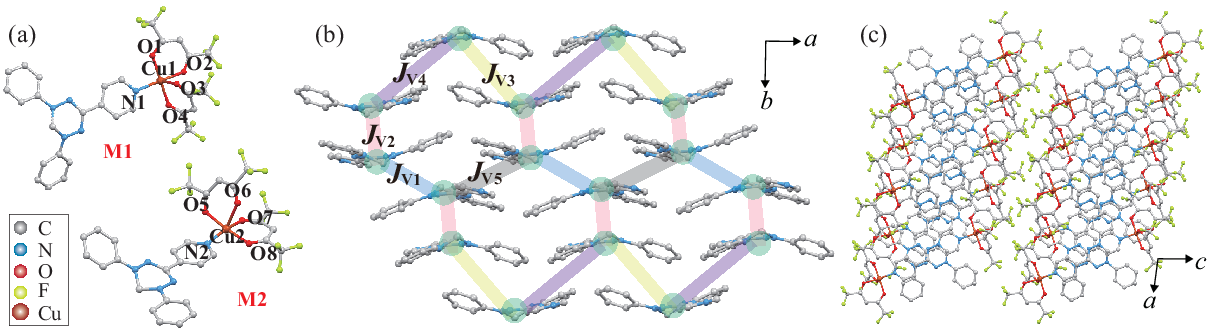}
\caption{(color online) (a) Two crystallographically independent molecules of ($p$-Py-V)[Cu(hfac)$_2$], which cause intramolecular exchange interactions of $J_{\rm{Cu1}}$ (M1) and $J_{\rm{Cu2}}$ (M2). 
The hydrogen atoms have been omitted for clarity. 
(b) Crystal structure forming a distorted honeycomb lattice composed of radicals in the $ab$ plane; each Cu(hfac)$_2$ in the molecule is omitted for clarity.
The green nodes represent the spin-1/2 of the radical. 
The thick lines represent the exchange interactions, $J_{\rm{V1}}$- $J_{\rm{V5}}$.
(c) Crystal structure viewed parallel to the honeycomb plane.
}
\end{center}
\end{figure*}

\begin{figure}[t]
\begin{center}
\includegraphics[width=18pc]{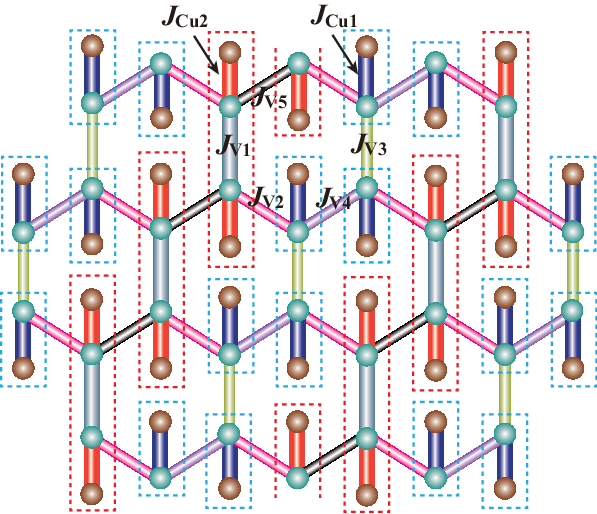}
\caption{(color online) Spin-1/2 distorted honeycomb-based lattice composed of intramolecular interactions, $J_{\rm{Cu1}}$ and $J_{\rm{Cu2}}$, and intermolecular interactions, $J_{\rm{V1}}$-$J_{\rm{V5}}$. 
Green and brown circles represent the spins of the racial and Cu atoms, respectively.
Only $J_{\rm{V5}}$ is ferromagnetic, whereas all others are AF, yielding frustration.
The broken lines enclose the $J_{\rm{Cu1}}$ dimers and $J_{\rm{Cu2}}$--$J_{\rm{V1}}$ tetramers.
}\label{f1}
\end{center}
\end{figure}

\begin{figure}[t]
\begin{center}
\includegraphics[width=20pc]{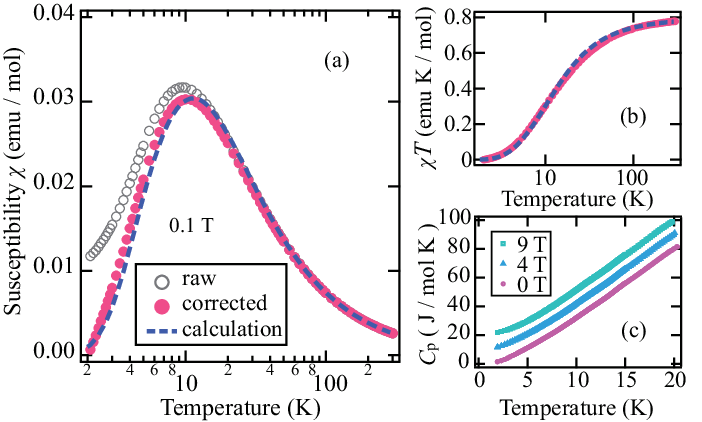}
\caption{(color online) Temperature dependence of (a) magnetic susceptibility ($\chi$ = $M/H$) and (b) $\chi T$ of ($p$-Py-V)[Cu(hfac)$_2$] at 0.1 T. 
The open circles denote raw data, and the closed circles are corrected for the paramagnetic term due to the impurity.
The broken lines represent the calculated results for the $J_{\rm{Cu1}}$ dimer and $J_{\rm{Cu2}}$--$J_{\rm{V1}}$ tetramer with $\alpha=J_{\rm{Cu1}}/J_{\rm{Cu2}}$ = 0.45 and $\beta =J_{\rm{V1}}/J_{\rm{Cu2}}$ = 0.40.
(c) Temperature dependence of the specific heat $C_{\rm{p}}$ of ($p$-Py-V)[Cu(hfac)$_2$] at 0, 4, and 9 T. 
For clarity, the values for 4 and 9 T have been shifted up by 10 and 20 J/ mol K, respectively.
}\label{f3}
\end{center}
\end{figure}

\begin{figure}[t]
\begin{center}
\includegraphics[width=20pc]{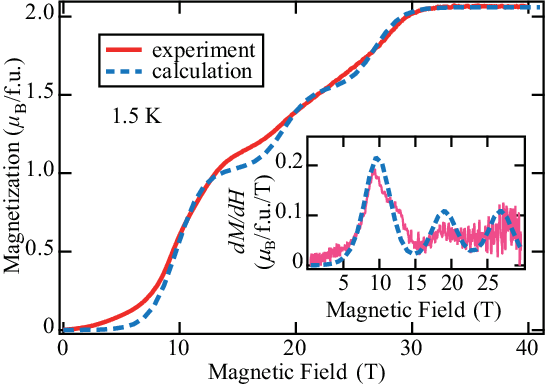}
\caption{(color online) Magnetization curve of ($p$-Py-V)[Cu(hfac)$_2$] at 1.5 K. 
The inset shows the field derivative of the magnetization curve.
The broken lines represent the calculated results for the $J_{\rm{Cu1}}$ dimer and $J_{\rm{Cu2}}$--$J_{\rm{V1}}$ tetramer with  $\alpha=J_{\rm{Cu1}}/J_{\rm{Cu2}}$ = 0.45 and $\beta =J_{\rm{V1}}/J_{\rm{Cu2}}$ = 0.40.
}\label{f3}
\end{center}
\end{figure}

\begin{figure}[t]
\begin{center}
\includegraphics[width=19pc]{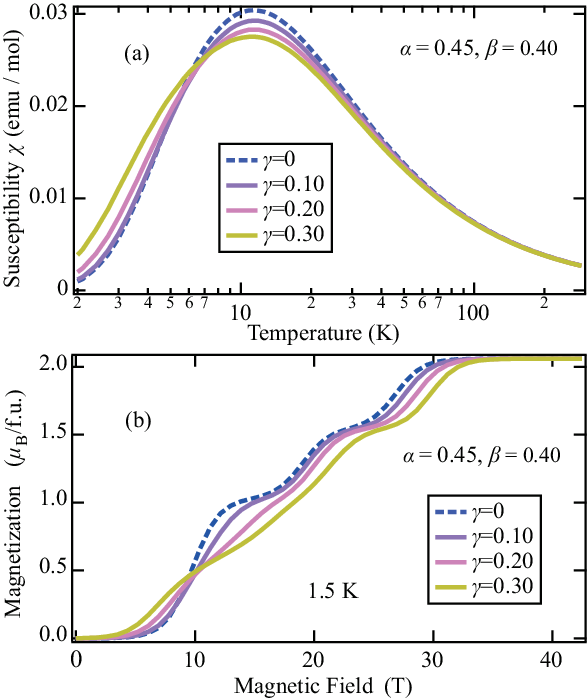}
\caption{(color online) (a) Calculated magnetic susceptibilities and (b) calculated magnetization curves at 1.5 K of the distorted honeycomb-based lattice assuming  $J_{\rm{V2}}$=$J_{\rm{V3}}$=$J_{\rm{V4}}$ and $J_{\rm{V5}}$=0 for the representative values of $\gamma=J_{\rm{V2}}/J_{\rm{Cu2}}$ with fixed $\alpha=J_{\rm{Cu1}}/J_{\rm{Cu2}}$ = 0.45 and $\beta =J_{\rm{V1}}/J_{\rm{Cu2}}$ = 0.40.  
}\label{f4}
\end{center}
\end{figure}

\subsection{Magnetic and thermodynamic properties}
In Figure 3(a), the temperature dependence of the magnetic susceptibility $\chi$ at 0.1 T is shown, indicating a broad peak at approximately 9.5 K. 
Figure 3(b) shows the temperature dependence of $\chi T$, which decreases as the temperature decreases, indicating dominant AF contributions.
In the low-temperature regime below the broad peak, $\chi$ exhibits a significant decrease, indicating the presence of a nonmagnetic ground state separated from the excited states by an energy gap.
Assuming the conventional paramagnetic behavior $C_{\rm{imp}}/T$ , where $C_{\rm{imp}}$ is the Curie constant of the spin-1/2 impurities, we evaluated the paramagnetic impurities to be $\sim$3.0 ${\%}$ of all spins, which is defined to fit the following calculated result. 

Figure 3(c) shows the temperature dependence of the specific heat.
We observe a monotonically decrease as the temperature decreases.
No sharp peak associated with a phase transition to an ordering is observed. 
Therefore, the specific heat is consistent with the gapped behavior observed in the magnetization curves shown below.
Schottky-like peak associated with the energy gap is masked by the lattice contributions of the verdazyl systems in the present temperature region~\cite{3D}.
Assuming a phase transition at lower temperatures, a clear upturn is expected to be observed even with lattice contributions.
Accordingly, the monotonic decreases demonstrate that the gapped state is stabilized up to 9 T.

Figure 4 presents the magnetization curve at 1.5 K under a pulsed magnetic field, revealing a sizable zero-field energy gap of approximately 10 T.
This gapped behavior is consistent with the $\chi$ behavior observed in the low-temperature region. 
Above 15 T, in the gapless region, the magnetization curve displays a two-step increase, which is clearly observed as two higher-field peaks in the field derivative of the magnetization curve ($dM/dH$), as shown in the inset of Fig. 4. 
Based on the isotropic $g$ value of 2.0 for organic radicals, a saturation value of 2.06 $\mu_{\rm{B}}$/f.u. suggests an average $g$ value of approximately 2.12 for the Cu spins.

\section{Analyses and Discussion}
The magnetic properties of the spin model were studied based on the results of MO calculations. 
Initially, we focused on the strong AF interactions of $J_{\rm{Cu1}}$, $J_{\rm{Cu2}}$, and $J_{\rm{V1}}$ to understand the the main characteristics of the spin model, which resulted in the formation of a dimer through $J_{\rm{Cu1}}$ and a tetramer composed of $J_{\rm{Cu2}}$ and $J_{\rm{V1}}$.
Using QMC method, we calculated the magnetic susceptibility and magnetization curves by considering the parameters $\alpha=J_{\rm{Cu1}}/J_{\rm{Cu2}}$ and $\beta =J_{\rm{V1}}/J_{\rm{Cu2}}$.
We obtained good agreement between the experimental and calculated results using $\alpha$ = 0.45 and $\beta$ = 0.40 ($J_{\rm{Cu2}}/k_{\rm{B}}$ = 30 K), as depicted in Figs. 3(a), 3(b), and 4.
The calculated results successfully reproduced the main features of the magnetization curve including the higher-field peaks observed in $dM/dH$.
It was confirmed that the quantum state in this model is primarily determined by three dominant AF interactions, forming dimers and tetramers.
The low-field gapped behavior was attributed to the nonmagnetic singlet state formed by the $J_{\rm{Cu1}}$ dimer, which exhibited a significant dependence on the parameter $\alpha$.
Subsequent bending of the magnetization curve at approximately 15 T corresponded to the full polarization of the spins forming the $J_{\rm{Cu1}}$ dimer, resulting in half-saturation magnetization. 
Above approximately 15 T, the two-step increase in the magnetization curve was attributed to the $J_{\rm{Cu2}}$--$J_{\rm{V1}}$ tetramer, where the ground state shifts to $S_{\rm{z}}$ = 0, $S_{\rm{z}}$ = 1, and $S_{\rm{z}}$ = 2 in the presence of a magnetic field. 
The region with $S_{\rm{z}}$ = 1 in the intermediate phase accounted for the higher-field two peaks in $dM/dH$, which strongly relied on the value of $\beta$.

Next, we investigated additional exchange interactions that contribute to the distorted honeycomb-based lattice.
To avoid the negative sign problem in the QMC calculation for frustrated systems, we assumed $J_{\rm{V5}}$ = 0.
Furthermore, the AF interactions $J_{\rm{V2}}$, $J_{\rm{V3}}$, and  $J_{\rm{V4}}$ were evaluated to be very close from the MO calculations, leading us to consider them equivalent ($J_{\rm{V2}}$=$J_{\rm{V3}}$=$J_{\rm{V4}}$) with $\gamma=J_{\rm{V2}}/J_{\rm{Cu2}}$ in the calculations. 
Figures 5(a) and 5(b) present the calculated magnetic susceptibility and magnetization curves as a function of $\gamma$, while keeping $\alpha$ = 0.45 and $\beta$ = 0.40 fixed. 
Increasing $\gamma$ that corresponds to the presence of 2D couplings tended to decrease the energy gaps originating from isolated states, resulting in a gradual decrease in the magnetic susceptibility below the broad peak temperature and the linearization of the magnetization curve. 
Nevertheless, it was evident that the contributions of $\gamma$, which represents the effects of 2D couplings without frustration, emphasize the difference between experimental and calculated results.
These finding suggest that frustration in the lattice of interest plays a crucial role in stabilizing the quantum-gapped state associated with the dominant AF interactions $J_{\rm{Cu1}}$, $J_{\rm{Cu2}}$, and $J_{\rm{V1}}$. 
In frustrated systems, dimensional reduction lowers the ground-state energy by lattice decoupling~\cite{tri1,tri2,tri3, PF6, Zn_gap}.
In the present lattice, the exchange interactions forming the 2D lattice are decoupled to minimize energy loss due to frustration, stabilizing the ground state characterized by the $J_{\rm{Cu1}}$ dimer and $J_{\rm{Cu2}}$--$J_{\rm{V1}}$ tetramer.
Furthermore, the ferromagnetic correlation also tends to stabilize the AF decoupled state~\cite{Zn_gap, hagiwara1}.

We observed several quantitative differences in the magnetization curves of the experimental results and calculations of the isolated quantum state ($\gamma$ = 0).
The discrepancies in the magnetization curve indicate that the actual energy state changes more continuously with a gentle slope when a magnetic field is applied, as shown in Fig. 4.
Moreover, the difference in the region between 10-15 T was prominent, where the observed $dM/dH$ exhibited a shoulder peak.
These differences may also be attributed to frustration effects in the present honeycomb-based lattice.
The Dzyaloshinskii-Moriya interactions that can exist between spins coupled by $J_{\rm{Cu1}}$, $J_{\rm{Cu2}}$, and $J_{\rm{V2}}$ are considered to be another origin of the gentle slope of the observed magnetization curve.

\section{Summary}
In this study, a verdazyl-based complex, ($p$-Py-V)[Cu(hfac)$_2$], was synthesized, resulting in a 5-coordinate environment with the Cu atom coordinated with the verdazyl ligand.
The MO calculations revealed five types of intermolecular interactions between the radical spins and two types of intramolecular interaction between the radical and Cu spins, leading to a spin-1/2 distorted honeycomb-based lattice.
The presence of competing ferromagnetic and AF interactions caused frustration.
The magnetic susceptibility exhibited dominant AF contributions and a gapped behavior. 
Consistently, no sharp peak associated with a phase transition to ordering was observed in the specific heat measurement. 
The magnetization curve exhibited a multistep increase, including a zero-field energy gap. 
By considering the stronger AF interactions forming a dimer and tetramer in the distorted honeycomb-based lattice, we could qualitatively explain the magnetic susceptibility and magnetization curves.
This analysis confirmed the stabilization of the quantum state based on the dominant AF interactions.
Furthermore, we demonstrated that the inclusion of 2D couplings without frustration enhanced the differences between the experimental and calculated results for the isolated quantum state.
These findings suggest that the exchange interactions forming the 2D lattice are decoupled to minimize the energy loss due to frustration, thereby stabilizing the quantum-gapped state.
This study proposes a spin model that showcases the synergistic effect of quantum fluctuation and frustration in honeycomb topology.
It is expected to inspire further research aimed at understanding quantum many-body phenomena arising from honeycomb lattices.

\begin{acknowledgments}
This research was partly supported by KAKENHI (Grants No. 23K13065 and No. 23H01127).
A part of this work was performed under the interuniversity cooperative research program of the joint-research program of ISSP, the University of Tokyo.
\end{acknowledgments}


\end{document}